\documentclass[epjCONF,columns]{svjour} 
\usepackage{graphics}
\usepackage{amsmath}
\usepackage[varg]{txfonts} 
\usepackage[latin1]{inputenc}
\session-title{
Hadron Collider Physics Symposium 2011, November 14-18, 2011, Paris, France}
\begin{document}
\title{Analysis strategy for the SM Higgs boson search in the four-lepton final state in CMS}
\author{Alberto Graziano\inst{1}\fnmsep\thanks{\email{alberto.graziano@cern.ch}} on behalf of the CMS Collaboration  
}
\institute{
Department of Experimental Physics, University of Torino/INFN, Torino, Italy
}
\abstract{
The current status of the searches for the SM Higgs boson in the $H$$\rightarrow$$ZZ^{(*)}$$\rightarrow$$4\ell$ decay channel with the CMS experiment~\cite{CMSexp} is presented. The selection cuts for suppressing the backgrounds while keeping very high signal efficiencies are described, along with the data-driven algorithms implemented to estimate the background yields and the systematic uncertainties. 
With an integrated luminosity of $1.66~\mathrm{fb}^{-1}$, upper limits at $95\%$ CL on the SM-like Higgs cross section  
$\times$
branching ratio exclude cross sections from about one to two times the expected value from the Standard Model in the range $150 < m_{H} < 420~\mathrm{GeV}$. No evidence for the existence of the SM Higgs boson has been found so far.
}
\maketitle

\section{Signal and backgrounds}
\label{sec:sig_bkg}
The \mbox{$H\rightarrow ZZ\rightarrow 4\ell$} analysis described here considers three final states ($4e$, $4\mu$, $2e2\mu$) over a Higgs boson mass range \mbox{$100 < m_{H} < 600~\mathrm{GeV/c^{2}}$}. 
The signature of signal events is very clean: two pairs of same-flavour, opposite-charge, high-$p_{T}$ isolated leptons pointing to the same reconstructed vertex are looked for. Mass constraints can be set on the dilepton invariant mass: also at low $m_{H}$, at least one Z boson is on-shell, therefore at least one pair of leptons has \mbox{$m_{\ell^{+}\ell^{-}} \simeq m_{Z}$}.

The $ZZ\rightarrow 4\ell$ background is often referred to as the \textit{irreducible} one, because of its signal-like kinematics. Since the SM Higgs boson is a scalar particle, angular correlations among the final-state leptons can be exploited to discriminate between the signal and the $ZZ$ background.

The \textit{reducible} backgrounds are $Zb\bar{b}$, $Zc\bar{c}$, $t\bar{t}$, with heavy-flavour quarks decaying semileptonically. 
Leptons originating from these decays usually point to displaced vertices and they are soft and  not isolated.
The \textit{instrumental} backgrounds are processes with jets misidentified as leptons, such as $Z+\mathrm{jets}$, $W+\mathrm{jets}$, QCD events.

\section{Event selection}
\label{sec:evt_sel}

\begin{figure}[h!]
\resizebox{0.95\columnwidth}{!}{
  \includegraphics{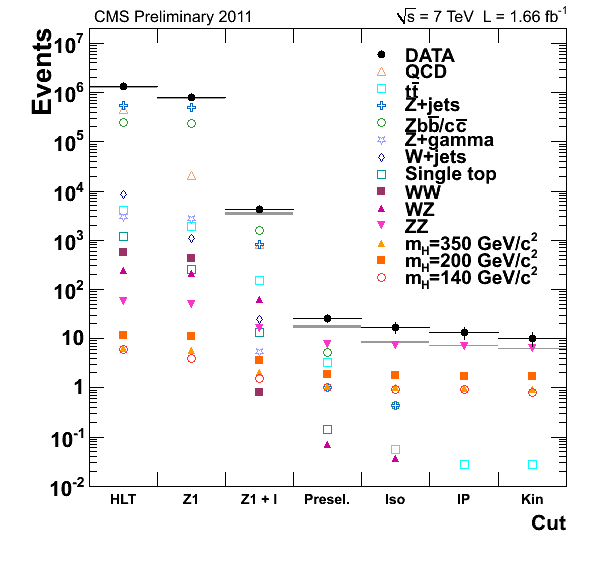} }
\caption{
Event yields in the $4\mu$ channel as a function of the event selection steps. Black points with uncertainties represent the data, other symbols represent the MC expectations. The samples correspond to an integrated luminosity of $L = 1.66~\mathrm{fb}^{-1}$.
}
\label{fig:step_plot}       
\end{figure}

The analysis consists of a set of cuts aiming to reduce the background contributions by preserving a high signal efficiency, as described in~\cite{PAS_2011_015}~\cite{PAS_2011_004}~\cite{AN_2011_123}. 

\begin{figure}[h!]
\resizebox{0.95\columnwidth}{!}{
  \includegraphics{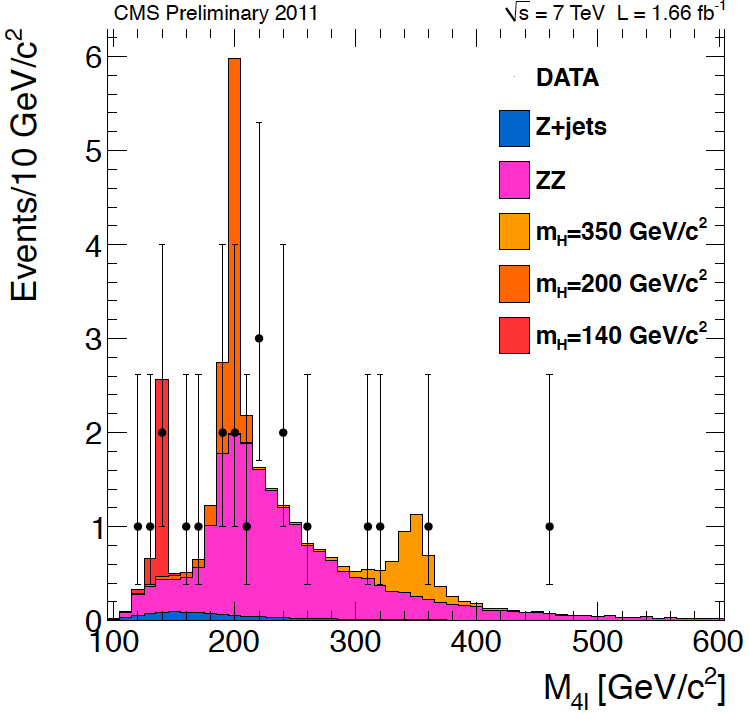} }
\caption{
Distribution of $m_{4\ell}$ after the baseline selection for the sum of the $4\ell$ channels. Points represent the data, shaded histograms represent the signal and background expectations. The samples correspond to an integrated luminosity of $L = 1.66~\mathrm{fb}^{-1}$.
}
\label{fig:baseline_final_4l}      
\end{figure}

\begin{itemize}
\item \textbf{Requirements on muons and electrons}\\
First of all, electrons and muons must satisfy some trigger, reconstruction and identification requirements. Depending on the \textit{trigger menu} deployed during data taking, reconstructed leptons are required to be matched to online trigger objects passing a single- or double-lepton trigger selection. They also have to pass loose $p_{T}$ and isolation cuts.
\\
\item \textbf{First Z candidate selection}\\
The first Z candidate is defined as the one with the dilepton invariant mass closest to the nominal Z mass, 
in the mass window \mbox{$60 < m_{\ell^{+}\ell^{-}} < 120~\mathrm{GeV/c^{2}}$},
after a selection including cuts on the $p_{T}$ of both leptons, on their isolation and on the significance of their 3D impact parameter with respect to the primary event vertex.
\\
\item \textbf{Z$_{1}+$ 1 lepton}\\
The presence of a third high-quality lepton of any flavour and charge is required. At this stage of the selection the phase space of the main reducible backgrounds is preserved for data-driven background estimation and control.
\\
\item \textbf{Z$_{1}+$ 2 leptons}\\
The requirement of a fourth lepton with matching flavour and opposite charge with respect to the third one is added.
\\
\item \textbf{\textit{`Best $4\ell$ candidate'} selection}\\
The second Z candidate is reconstructed from the two highest-$p_{T}$ leptons not associated to $Z_{1}$ and passing \mbox{$m_{Z_{2}} > 12~\mathrm{GeV/c^{2}}$}. 
At this stage the ambiguity due to combinatorics in events with extra fake leptons is limited and the \textit{`best $4\ell$ candidate'} is chosen.

The $4\ell$ candidate must satisfy \mbox{$m_{4\ell} > 100~\mathrm{GeV/c^{2}}$}. Moreover, in the $4e$ and $4\mu$ final states only, at least three out of the four possible $\ell^{+}\ell^{-}$ combinations are required to have $m_{\ell^{+}\ell^{-}} > 12~\mathrm{GeV/c^{2}}$ in order to reject background events with leptonic $J/\psi$ decays.
\\
\item \textbf{Cut on relative lepton isolation}\\
The two leptons with the largest isolation variable, which is the sum of tracker, ECAL and HCAL isolation divided by the lepton $p_{T}$, are then considered.
The sum of their isolation values is required to be lower than a threshold.
\\
\item \textbf{Cut on the 3D impact parameter significance of leptons}\\
The displaced vertex of leptons originating from $b$-quark decays can be a handle for further background rejection.
A cut is therefore applied on the significance of the 3D impact parameter of the lepton track with respect to the reconstructed primary vertex. This significance is defined as $SIP_{3D} = IP_{3D}/\sigma_{IP_{3D}}$ and its absolute value is required to be less than 4 for all selected leptons.
\\
\item \textbf{Cut on Z$_{1}$, Z$_{2}$ kinematics}\\
Finally, additional constraints are imposed on the $p_{T}$ of the selected leptons 
(\mbox{$p_{T}^{\ell_{1},\ell_{2},\mu_{3},\mu_{4}} > 20, 10, 5, 5~\mathrm{GeV/c}$} for muons, 
\mbox{$p_{T}^{\ell_{1},\ell_{2},e_{3},e_{4}} > 20, 10, 7, 7~\mathrm{GeV/c}$} for electrons)
and on the invariant mass of the second $Z$ candidate, which is required to be \mbox{$20 < m_{Z_{2}} < 120~\mathrm{GeV/c^{2}}$} in the \textit{baseline} selection and \mbox{$60 < m_{Z_{2}} < 120~\mathrm{GeV/c^{2}}$} in the \textit{high mass} selection. 
The $m_{4\ell}$ distribution for events passing the baseline selection is shown in Fig.~\ref{fig:baseline_final_4l}.
\end{itemize}

\section{$ZZ\rightarrow 4\ell$ cross section measurement}
\label{sec:ZZ_xsec}
The $ZZ\rightarrow 4\ell$ inclusive cross section has been measured after the \textit{high mass} selection cuts \mbox{$60<m_{Z_{1}}<120~\mathrm{GeV/c^{2}}$}, 
\mbox{$60<M_{Z_{2}} < 120~\mathrm{GeV/c^{2}}$} as
\begin{equation}
\begin{split}
\sigma(&pp\rightarrow ZZ+X) \times BR(ZZ\rightarrow4\ell) = \frac{\displaystyle\sum (N_{obs}^{i_{ch}} - N_{bkg}^{i_{ch}})}{\displaystyle \mathcal{A}_{4\ell} \times \varepsilon_{ZZ\rightarrow4\ell} \times \mathcal{L}} = \\
&= 20.84_{-4.0}^{+6.8}~\mathrm{(stat.)}~\pm~0.54\mathrm{(syst.)}~\pm~0.94\mathrm{(lumi.)}~\mathrm{fb}
\end{split}
\end{equation}
where the sum runs over the three final states ($4e$, $4\mu$, $2e2\mu$).
This result should be compared with the theoretical value:
\begin{center}
$\sigma_{TH}(pp\rightarrow ZZ+X) \times BR(ZZ\rightarrow4\ell)~=~28.32~\pm~2.57~\mathrm{fb}$
\end{center}

\section{ZZ background control}
\label{sec:ZZ_bkg}
The contribution from ZZ background, which is the main one after the whole event selection, can be estimated from the number of $Z\rightarrow 2\ell$ events observed in data with the following formula:
\begin{equation}
N_{ZZ\rightarrow 4\ell} = \frac{\sigma_{q\bar{q}\rightarrow ZZ\rightarrow 4\ell}^{NLO}+\sigma_{gg\rightarrow ZZ\rightarrow 4\ell}^{LO}}{\sigma_{pp\rightarrow Z\rightarrow 2\ell}^{NNLO}} \cdot \frac{\varepsilon_{ZZ\rightarrow 4\ell}^{MC}}{\varepsilon_{Z\rightarrow 2\ell}^{MC}} \cdot N_{Z\rightarrow 2\ell}^{observed}
\end{equation}
This method exploits the fact that most Feynman diagrams are shared by the two processes. The results are compatible with those obtained directly from MC, but the systematic uncertainties are smaller because most of them cancel out in the ratio.

\section{$Zb\bar{b}$, $Zc\bar{c}$, $t\bar{t}$ background control}
\label{sec:Zbb_tt_bkg}
\begin{figure}[h!]
\resizebox{0.90\columnwidth}{!}{
  \includegraphics{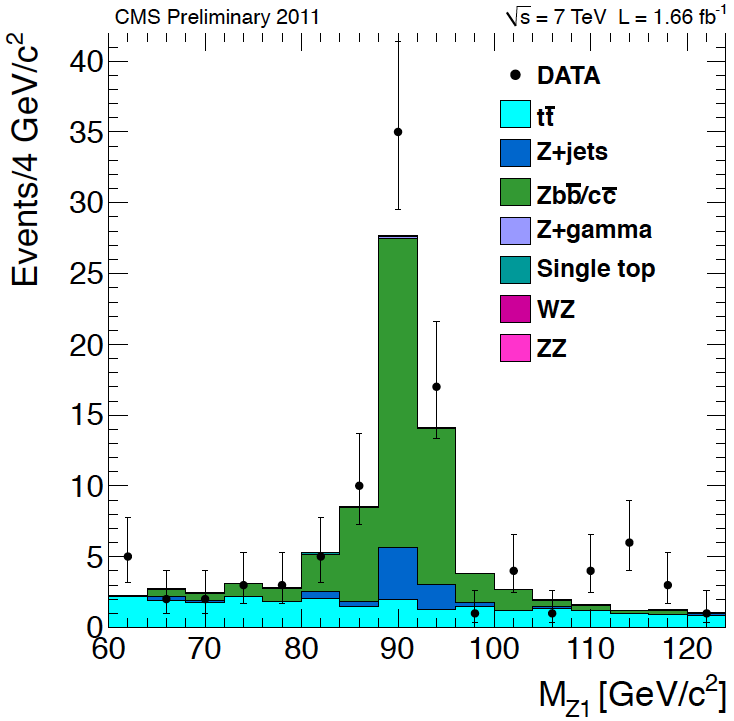} }
\caption{
Distribution of the best reconstructed Z candidate invariant mass for the events in the $4\ell$ background control region. Solid points represent the data, shaded histograms represent the MC expectations. The signal and the ZZ background contribute negligibly. The samples correspond to an integrated luminosity of $L = 1.66~\mathrm{fb}^{-1}$.
}
\label{fig:Zbb_tt}   
\end{figure}
In order to perform a data-driven measurement of the $Zb\bar{b}$, $Zc\bar{c}$, $t\bar{t}$ background yield, 
the first $Z$ candidate is defined as in the signal selection, whereas the flavour, charge and isolation requirements on the leptons from $Z_{2}$ are relaxed and the cut on their impact parameter significance reversed: $|SIP_{3D}| > 5$
(see Fig.~\ref{fig:Zbb_tt}).

To propagate the event yield from this control region to the signal phase space, correction factors are introduced that account for the relaxed isolation and kinematical cuts, for the reversed impact parameter cut and for the combinatorics related to considering pairs of leptons of any flavour and charge.

\section{$Z+\mathrm{jets}$ background control}
\label{sec:Zjets_bkg}

\subsection{Single lepton fake rate measurement}
Prior to measuring the $Z+$jets background yield, the \textit{single lepton fake rate} must be evaluated. This is done from a sample of exactly 3 leptons, therefore signal-free, in which the contamination from $WZ$ events is removed with a cut on the missing transverse energy ($\ensuremath{{\not\mathrel{E}}_T} < 25~\mathrm{GeV/c^{2}}$). A $Z_{1}$ candidate is looked for like in the signal selection and the remaining lepton (a \textit{`fakeable object'}) is required to pass identification and isolation cuts. The fraction of fakeable objects passing this selection, as a function of lepton $p_{T}$ and pseudorapidity,
is the single lepton fake rate:
\begin{equation}
\varepsilon(p_{T}^{\ell}, \eta^{\ell}) = \frac{\mathrm{N(passing~ID~and~isolation~cuts)}}{\mathrm{N(fakeable~objects)}}
\end{equation}

\subsection{Definition of the control region}
A $Z_{1}$ candidate is reconstructed as in the signal selection.
The control region is signal-free because the third and fourth leptons are required to have the same flavour and charge, $\ell_{3}^{\pm}\ell_{4}^{\pm}$ (the lepton charge misassignment is negligible).
No identification and isolation cuts are applied on these two leptons, whereas the kinematical cuts \mbox{$m_{\ell_{3}\ell_{4}} > 12~\mathrm{GeV/c^{2}}$}, \mbox{$m_{4\ell} > 100~\mathrm{GeV/c^{2}}$}
are.

\subsection{Extrapolation to the signal region}
\begin{figure}[h!]
\resizebox{0.50\columnwidth}{!}{
  \includegraphics{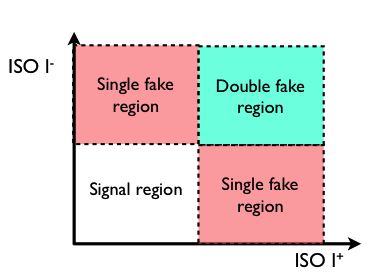} }
\resizebox{0.49\columnwidth}{!}{
  \includegraphics{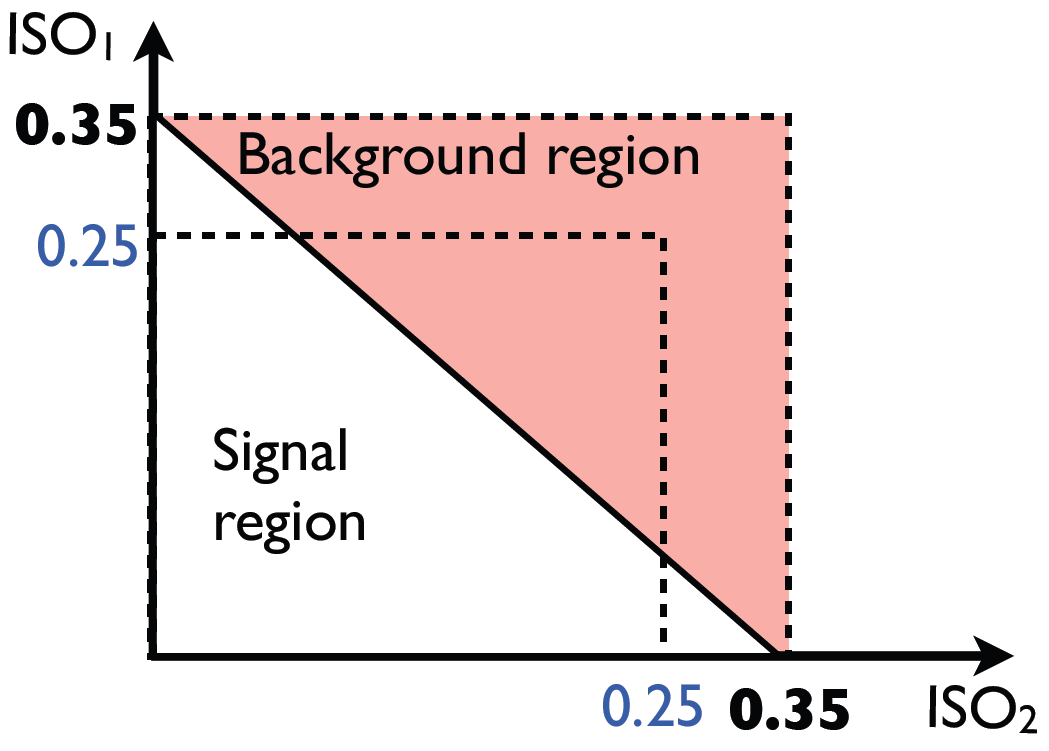} }
\caption{
Two sketches about the extrapolation from the control region to the signal one for the evaluation of the $Z+\mathrm{jets}$ event yield. Left: the CR, called `Double fake region', is shown in green, the SR in white. Right: the SR is defined by 
$ISO_{\ell_{3}}+ISO_{\ell_{4}} < 0.35$.
}
\label{fig:Zjets_sketch1}
\end{figure}

The number of $Z+$jets events in the signal region (SR) can be extrapolated from the one in the control region (CR) by means of the following formula:
\begin{equation}
N_{SR}^{Z+jets} = N_{CR}^{Z+jets} \times \frac{1}{2}~\times~\frac{\varepsilon(p_{T}^{\ell_{3}}, \eta^{\ell_{3}}) \times \varepsilon(p_{T}^{\ell_{4}}, \eta^{\ell_{4}})}{[1-\varepsilon(p_{T}^{\ell_{3}}, \eta^{\ell_{3}})]~\times~[1-\varepsilon(p_{T}^{\ell_{4}}, \eta^{\ell_{4}})]}
\end{equation}
where $N_{CR}^{Z+jets}$ is scaled by the ratio of the joint probability of both $\ell_{3}$ and $\ell_{4}$ passing the ID and isolation selection (as expected in the SR) to the joint probability of neither of them passing it (as expected in the CR).
The factor $1/2$ accounts for the triangular area of the signal region in the ($ISO_{\ell_{3}}$, $ISO_{\ell_{4}}$) plane. See Fig.~\ref{fig:Zjets_sketch1}.

\section{Exclusion limits for $\sqrt{s}=7~\mathrm{TeV}$, $L=1.66~\mathrm{fb}^{-1}$}
\label{sec:excl_limit}

\begin{figure}[h!]
\resizebox{0.95\columnwidth}{!}{
  \includegraphics{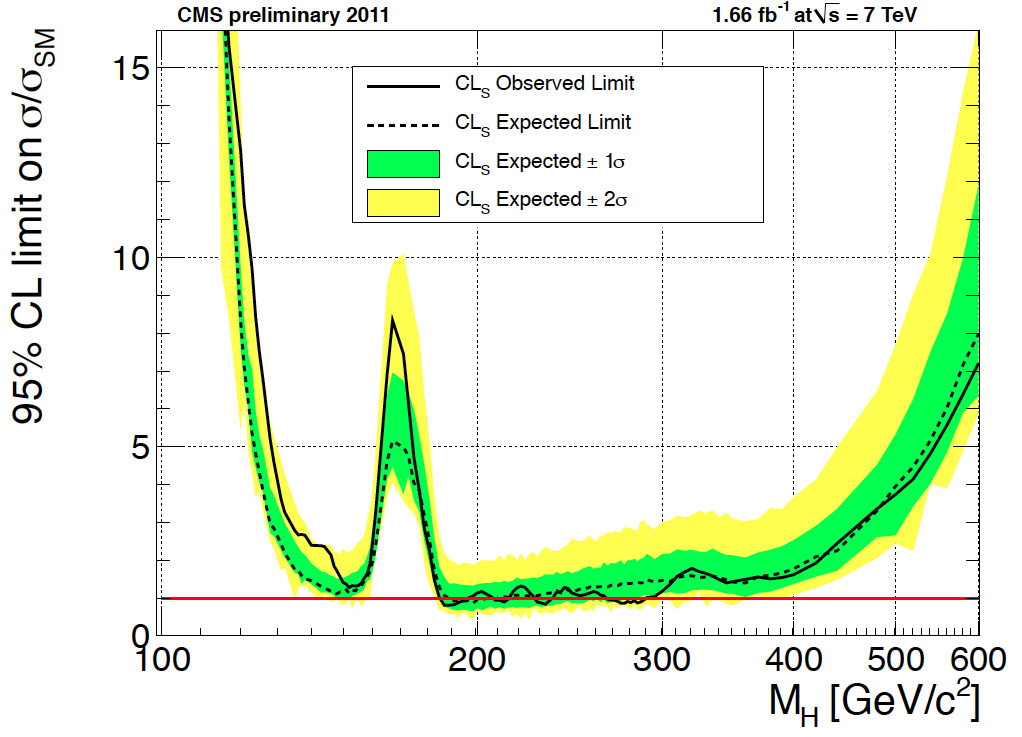} }
\caption{
The mean expected and the observed upper limits at $95\%$ CL on \mbox{$\sigma(pp\rightarrow H+X) \times BR(ZZ\rightarrow 4\ell)$} for a Higgs boson in the mass range $120\div600~\mathrm{GeV/c^{2}}$, for an integrated luminosity of $1.66~\mathrm{fb}^{-1}$ using the $CL_{s}$ approach. The expected ratio for the SM is presented. The results are obtained using a shape analysis method.
}
\label{fig:exclusion}     
\end{figure}

With $\sqrt{s} = 7~\mathrm{TeV}$ and $L=1.66~\mathrm{fb}^{-1}$, the search for a SM-like Higgs boson performed in the \mbox{$H\rightarrow ZZ^{(*)}\rightarrow 4\ell$} channel allows to set $95\%$ CL upper limits on \mbox{$\sigma \cdot BR$} that exclude cross sections from about one to two times the expected SM ones in the mass range
\mbox{$150 < m_{H} < 420~\mathrm{GeV/c^{2}}$}. See Fig.~\ref{fig:exclusion}.

%
%

\end{document}